\documentclass{jps-cp}
\usepackage{txfonts} 
\usepackage{rotating}
\usepackage{graphicx}
\usepackage{dcolumn}
\usepackage{amstext}
\usepackage{lscape}
\usepackage{verbatim}
\title{Neutron skins of heavy nuclei and tidal deformability of neutron star}

\author{Bharat \textsc{Kumar}$^{1,2}$ }

\inst{$^{1}$Center for Computational Sciences, University of Tsukuba, Tsukuba 305-8577, Japan \\
$^{2}$Inter-University Centre for Astronomy and Astrophysics, Pune - 411007, India}

\email{bharat@nucl.ph.tsukuba.ac.jp}

\recdate{August 8, 2019}

\abst{In this paper, I have discussed the numerical predictions for the neutron-skin thickness (NST) of various finite nuclei starting from $^{40}$Ca to $^{238}$U using recently developed effective relativistic mean-field models G3 and IOPB-I \cite{G3, IOPB}.  The calculated results are compared with the PREX-II data, and the experiment has been done with antiprotons at CERN. Further, I have also calculated the dimensional tidal deformability of a canonical neutron star 1.4$M_\odot$  and compared it with the recent observation of GW1701817. }
\kword{Neutron-skin thickness, Neutron star, Gravitational waves}

\begin{document}
\maketitle

\section{Introduction}
The accurate description of the matter distribution in nuclei is an important problem in nuclear physics to understand the nuclear structure of the nuclei.  The proton distribution has been measured with very high accuracy using electron-nucleus elastic scattering. But it is more difficult to accurately determine the neutron density distributions of nuclei by any experimental probe because of the electrically neutral particle.  Several experiments of charge distribution of neutron have been carried out worldwide over the last decade by using proton scattering and interaction cross-sections in heavy-ion collisions at relativistic energies. Till now,  the determination of neutron radii is very poorly known compared to proton radii. However, the newly designed experiments such as PREX-II at JLab and the Bates Laboratory at MIT with polarized beams and targets have resulted in a significant better clearer picture of the neutron charge distribution \cite{BATES, PREX}.  However, PREX-II experiment has mainly designed for neutron radius of $^{208}$Pb, and give the first model-independent neutron-skin thickness (NST) of $^{208}$Pb to be $0.33^{+0.16}_{-0.18}$ fm.   The total error is almost half the median value, but the PREX  experiment is exciting, and future higher statistics data are required to reduce the uncertainty. Using the tight bound of the tidal deformability of the canonical neutron star, Fattoyev {\it et al.} have suggested an upper limit of NST of $^{208}$Pb to be about $\Delta r_{np}\lesssim0.25$ fm \cite{SKIN}.  

Thanks to GW170817, the first detected binary neutron star merger. The first analysis of the tidal phasing has placed the upper bound of  $\Lambda \le 800$  at 90$\%$ confidence on the tidal deformability of a $1.4M_{\odot}$ neutron star \cite{BNS}. This analysis did not explicitly consider that both compact objects were neutron stars. Then they have re-analysed again using an ordinary equation of state for both the stars and placed $\Lambda_{1.4} = 190^{+390}_{ -120}$ that translates to a stringent bound of $\Lambda_{1.4} =580$ at the 90$\%$ confidence level \cite{BNS1}. These bounds can, in turn, constrain the neutron star equation of state. 

In this contribution, I have calculated NST for the finite nuclei starting from $^{40}$Ca to $^{238}$U using the recently developed new energy density functional G3 and IOPB-I.  Next, I have shown the result of the dimensional tidal deformability of the neutron stars, which was observed by the aLIGO-VIRGO collaboration \cite{BNS, BNS1}. 

Throughout this paper, I have used geometric units, c=1=G, where c is the speed of light, and G is the gravitational constant, respectively. 

\section{The extended relativistic mean-field (ERMF) model}\label{sec2}
The relativistic mean-field model is the effective model of QCD formalism. The mean-field formalism approximates the effect of vacuum fluctuation and Fock term in the calculation by fitting the coupling constant with experimental observables. It is the virtue of mean-field that can ignore the basic formalism difficulties like renormalization and divergence of the system and take the simple way just by fitting the coupling constant. In this calculation, I have taken the non-linear coupling in the $\sigma-$meson like Boguta Lagrangian, $\omega-\rho$ cross-coupling. Additionally, I have included the $\delta-$meson, that is responsible for the mass isospin asymmetry of the nucleon.  Here,  I am starting with the energy density functional for the ERMF model \cite{G3, IOPB}: 

\begin{eqnarray}
{\cal E}({r})&=&  \sum_\alpha \varphi_\alpha^\dagger({r})
\Bigg\{ -i \mbox{\boldmath$\alpha$} \!\cdot\! \mbox{\boldmath$\nabla$}
+ \beta \left[M - \Phi (r) - \tau_3 D(r)\right]\nonumber 
+ W({r})+ \frac{1}{2}\tau_3 R({r}) + \frac{1+\tau_3}{2} A ({r})\nonumber \\
& &
- \frac{i \beta\mbox{\boldmath$\alpha$}}{2M}\!\cdot\!
  \left (f_\omega \mbox{\boldmath$\nabla$} W({r})
  + \frac{1}{2}f_\rho\tau_3 \mbox{\boldmath$\nabla$} R({r})+f \mbox{\boldmath$\nabla$} A
 \right)
+\frac{1}{2 M^2} (\beta_\sigma+\beta_\omega \tau_3)\Delta A 
  \Bigg\} \varphi_\alpha (r)\nonumber\\
&& + \left ( \frac{1}{2}
  + \frac{\kappa_3}{3!}\frac{\Phi({r})}{M}
  + \frac{\kappa_4}{4!}\frac{\Phi^2({r})}{M^2}\right )
   \frac{m_s^2}{g_s^2} \Phi^2({r})
-  \frac{\zeta_0}{4!} \frac{1}{ g_\omega^2 } W^4 ({r})
+ \frac{1}{2g_s^2}\left( 1 +
\alpha_1\frac{\Phi({r})}{M}\right) \left(
\mbox{\boldmath $\nabla$}\Phi({r})\right)^2
 \nonumber \\ 
&&- \frac{1}{2g_\omega^2}\left( 1 +\alpha_2\frac{\Phi({r})}{M}\right)
\left( \mbox{\boldmath $\nabla$} W({r})  \right)^2 
- \frac{1}{2}\left(1 + \eta_1 \frac{\Phi({r})}{M} +
 \frac{\eta_2}{2} \frac{\Phi^2 ({r})}{M^2} \right)
  \frac{m_\omega^2}{g_\omega^2} W^2 ({r})
- \frac{1}{2e^2} \left( \mbox{\boldmath $\nabla$} A({r})\right)^2 \nonumber \\
&& - \frac{1}{2g_\rho^2} \left( \mbox{\boldmath $\nabla$} R({r})\right)^2
- \frac{1}{2} \left( 1 + \eta_\rho \frac{\Phi({r})}{M} \right)
   \frac{m_\rho^2}{g_\rho^2} R^2({r})
- \frac{\eta_{2\rho}}{4 M^2}\frac{{m_\rho}^2}{{g_\rho}^2}\left(R^{2}(r)\times W^{2}(r)\right) \nonumber \\
&&+\frac{1}{2 g_{\delta}^{2}}\left( \mbox{\boldmath $\nabla$} D({r})\right)^2
   +\frac{1}{2}\frac{ {  m_{\delta}}^2}{g_{\delta}^{2}}\left(D^{2}(r)\right)
-\frac{1}{2 e^2}(\mbox{\boldmath $\nabla$}A)^2+\frac{1}{3 g_\gamma g_\omega}
A \Delta W +\frac{1}{g_\gamma g_\rho}A\Delta R,\;
\label{eq1}
\end{eqnarray}
where, $\Phi, W, R, D$ and $A$ are the fields and $g_{\sigma},g_{\omega},g_{\rho},g_{\delta}$, and $\frac{e^2}{4\pi}$ are the coupling constants of the fields. $m_{\sigma},m_{\omega},m_{\rho}$ and $m_{\delta}$ are the masses of the mesons for $\sigma,\omega,\rho,\delta$ and photon field, respectively. The extended energy density functional with  $\delta-$meson contains the nucleons and other exchange mesons like $\sigma$, $\omega$ and $\rho-$meson and photon $A_{\mu}$. The effects of the $\delta-$meson to the bulk properties of finite nuclei are nominal, but the effects are significant for highly asymmetric dense nuclear matter. The $\delta-$ meson causes the splits of the effective masses of proton and neutron which modify the properties of the neutron star and heavy-ion reaction. Additionally, from the energy density functional in Eq.\ref{eq1}, the terms having $g_\gamma, f,  \beta_\sigma $ and $\beta_\omega$ are responsible for the effects related to the electromagnetic structure of the pion and nucleon \cite{furnstahl97}.

\section{Results}
\begin{figure}[h]
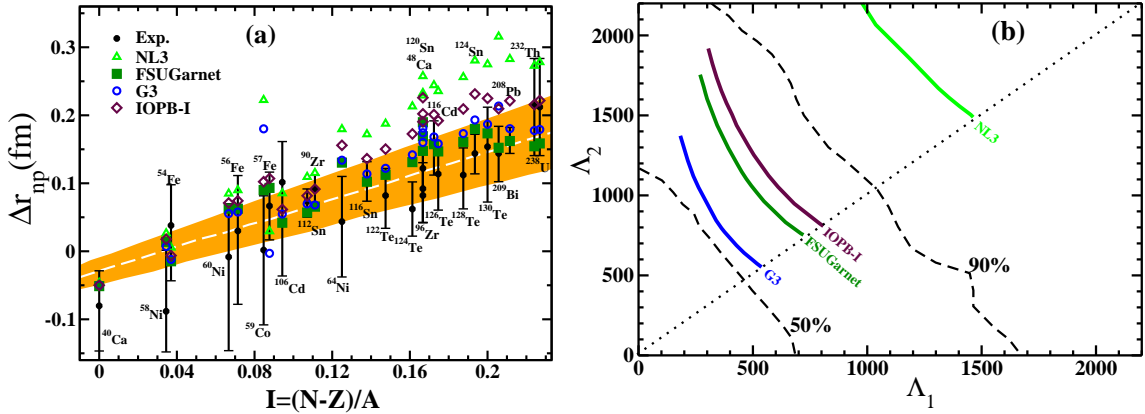

    \centering
    \includegraphics[width=0.47\textwidth]{skin.eps}
    \includegraphics[width=0.495\textwidth]{tidal.eps}
\caption{(a) The NST as a function of the asymmetry parameter of the nucleus $I=(N-Z)/A$. Results obtained with the parameter sets IOPB-I, and G3 are compared with experimental values \cite{thick}. (b) Tidal deformability parameters for the case of high mass ($\Lambda_1$) and low-mass ($\Lambda_2$) components of the observed GW170817. The figures are adopted from \cite{IOPB}.}
\label{fig1}    
\end{figure}

I have calibrated the parameters of the energy density functional as given by Eq.(1). The optimization of the energy density functional performed for a given set of fit data using the simulated annealing method. This method allows one to search for the best fit parameter in a given region of the parameter space.  I have fitted the coupling constants of Eq.(1) to the properties of eight spherical nuclei together with some constraints on the properties of the nuclear matter at the saturation density.  Newly generated ERMF models G3 and IOPB-I are very well suitable for the study ground-state properties of finite nuclei, nuclear matter, and neutron star.  Here,  I have examined the NST of  26 stable nuclei starting from $^{40}$Ca and $^{238}$U without considering the nuclear deformation and superfluidity \cite{thick}. The result for the difference between the neutron and proton rms radii gives the NST of the nucleus. The NST $\Delta r_{np}$ is defined as
\begin{eqnarray}
\Delta r_{np} = R_n - R_p.
\end{eqnarray}
Where $R_n$ and $R_p$ are the rms radii for the neutron and proton distribution respectively. The addition of $\omega-\rho$ cross-coupling into the Lagrangian density constrains the NST of finite nuclei. Fig.\ref{fig1}(a) shows the NST $\Delta r_{np}$ for $^{40}$Ca to $^{238}$U nuclei as a function of proton-neutron asymmetry $I=(N-Z)/A$. The calculated results of $\Delta r_{np}$ for NL3, FSUGarnet, G3, and IOPB-I parameter sets compared with the experimental results deduced from the antiprotons at CERN \cite{thick}. The experimental data with the error bars shown for 26 stable nuclei ranging from $^{40}$Ca to $^{238}$U. The shaded region (orange) is the fitted experimental data which shows an approximately linear dependence of NST on the relative neutron excess of a nucleus. The results of the parameter set G3 and IOPB-I  are also shown in Fig. \ref{fig1}(a). The values of the NST of IOPB-I set for some of the nuclei slightly deviate from the shaded region, because IOPB-I has a smaller strength of $\omega-\rho$ cross-coupling as compared to the FSUGarnet set. However, the G3 set is almost compatible with the experimental data.  The calculated values of NST for the $^{208}$Pb nucleus are 0.283, 0.162, 0.180, and 0.221 fm for the NL3, FSUGarnet, G3, and IOPB-I parameter sets, respectively.  The calculated values of $\Delta r_{np}$ for G3 and IOPB-I are consistent with the upper limit of $\Delta r_{np}\le 0.25$ fm, which obtained with the help of correlation between the NST and with canonical tidal deformability of the neutron star \cite{SKIN}.

Finally, in this section, I will discuss the neutron star properties such as the tidal deformability of the binary neutron star.  To calculate the tidal deformability, I am solving the TOV equation together with the perturbing metric, where the nuclear equation of state use as an input. Tidal deformability is highly sensitive to the fifth power of the radius ($R^5$) or the compactness ($C=M/R$) of the star. The relation defines the dimensionless version of the tidal deformability \cite{tanja}: 
\begin{equation}
\Lambda = \frac{\lambda}{M^5}=\frac{2 k_2}{3 C^5}
\end{equation}
where $k_2$ is the dimensionless tidal love number which gives the internal information of the neutron star. $M$ is the mass of the star. 
For the binary neutron star, the  individual quantities of the tidal deformability $\Lambda_1$ and $\Lambda_2$ associated with the phase of the gravitational wave; which is given by
\begin{equation}
 \widetilde{\Lambda} \!=\! \frac{16}{13} \!\left[
   \frac{(M_{1}\!+\!12M_{2})M_{1}^{4}}{(M_{1}\!+\!M_{2})^{5}}\Lambda_{1} \!+\!
   \frac{(M_{2}\!+\!12M_{1})M_{2}^{4}}{(M_{1}\!+\!M_{2})^{5}}\Lambda_{2} \right].
   \label{LambdaTilde}
\end{equation}
It is noticed that $\widetilde{\Lambda}\!=\!\Lambda_{1}\!=\!\Lambda_{2}$ is for the equal-mass neutron star. Notably, the tidal deformability determined from the first BNS merger is already convincing enough to rule out the various types of viable EoSs. Fig. \ref{fig1}(b)  shown the $\Lambda_1$ and $\Lambda_2$ values for the fixed chirp mass $1.188M_{\odot}$ from GW170817 for the high-mass $M_1$, and low-mass $M_2$ components of the binary system. From the GW170817, the values of $\Lambda\leq800$ (first analysis) and $\Lambda=190^{+390}_{-120}$ (re-analysis the GW170817 data) in the low-spin case are within the $90\%$ credible intervals which are consistent with the 680.79, 622.06, and 461.03 of the $1.4M_\odot$ NS binary for the IOPB-I, FSUGarnet, and G3 parameter sets, respectively . Furthermore, the NL3 equation of state is ruled out on the basis of the GW170817 data  because of the stiffer nature. 

\section{Conclusion}
In this analysis, I have studied the NST of various finite nuclei and tidal deformability of the neutron star using the effective interactions G3 and IOPB-I. Both parameter sets have very well reproduced the finite nuclei, infinite nuclear matter, and neutron star properties. In particular, the calculated value $\Lambda_{1.4}$ of a neutron star is consistent with the recent observation of the GW170817. \\

{\bf Acknowledgement:}This work is supported in part by JSPS KAKENHI Grant No.18H01209.

\end{document}